\documentclass[a4paper,twocolumn,american,aps]{revtex4}
\usepackage[T1]{fontenc}
\usepackage[latin1]{inputenc}
\usepackage{subfigure}
\usepackage{amsmath}
\usepackage{graphicx}
\usepackage{amssymb}

\makeatletter

\providecommand{\tabularnewline}{\\}

\usepackage{float}
\usepackage{graphicx}

\usepackage{babel}
\makeatother

\begin{document}
\begin{abstract}
We calculate the spectral functions of model systems describing $5f$-compounds
adopting Cluster Perturbation Theory. The method allows for an accurate
treatment of the short-range correlations. The calculated excitation
spectra exhibit coherent $5f$ bands coexisting with features associated
with local intra-atomic transitions. The findings provide a microscopic
basis for partial localization. Results are presented for linear chains.
\end{abstract}

\title{Spectral functions for strongly correlated $\textrm{5f}$-electrons}

\author{F. Pollmann and G. Zwicknagl}

\address{\emph{Institut f\"ur Mathematische Physik, Technische Universit\"at Braunschweig,
Mendelsohnstr. 3, 38106 Braunschweig, Germany}}

\maketitle

\section{Introduction}

Electronic correlations are strongly evident in actinide intermetallic
compounds. The highly complex phase diagrams with novel sometimes
enigmatic ordered states reflect the sensitivity to small changes
in external control parameters like temperature, pressure, magnetic
fields \cite{Thalmeier05}. An undispensible prerequisite for an explanation
of the observed anomalies is a microscopic understanding of the strongly
interacting $5f$ electrons which partially preserve atomic-like character.
Occupying the partially filled $5f$ shells according to Hund's rules
leads to magnetic moments. The lifting of rotational symmetry in a
crystal by the crystalline electric field (CEF) and the hybridization
with the delocalized conduction states of the outer shell electrons
leads to a large number of low-energy excitations. 

Of particular interest are the heavy fermion phases where the low-energy
excitations correspond to heavy quasi-particles. Their enhanced effective
masses $m^{*}$ are reflected in enhanced values of the Sommerfeld
coefficient and the Pauli spin susceptibility. The heavy quasi-particles
have been observed by de Haas-van Alphen (dH-vA) experiments in a
number of compounds. The experiments unambiguously confirm that some
of the U $5f$ electrons must have itinerant character. It has been
known for quite some time that the $5f$-states in actinide intermetallic
compounds cannot be considered as ordinary band states. Standard band
structure calculations based on the Local Density Approximation (LDA)
fail to reproduce the narrow quasiparticle bands. On the other hand,
the predicted bandwidths are too small to explain photoemission data
\cite{Allen92,Fujimori99}. These shortcomings are a direct consequence
of the inadequate treatment of local correlations within ordinary
electron structure calculation.

Increasing experimental evidence points towards a two-fluid model
implying the co-existence of both localized atomic-like and itinerant
band-like $5f$ states. For U heavy fermion compounds the photoemission
spectra usually display a two-peaked structure in the $5f$ emission
\cite{Arko87,Eloirdi05}. The peak at $1$ eV binding energy results
from the localized $5f$ states while the narrow peak at the Fermi
level is attributed to $5f$ derived itinerant quasi-particles \cite{Kumigashira00}.
Concerning the low-energy excitations it has been shown recently that
the dual model allows for a quantitative description of the renormalized
quasi-particles - the heavy fermions - in UPd$_{2}$Al$_{3}$. The
measured dH-vA frequencies for the heavy quasiparticle portions as
well as the large anisotropic effective masses can be explained very
well by treating two of the $5f$ electrons as localized \cite{Zwicknagl03a,Zwicknagl03}.
Finally, the co-existence of $5f$-derived quasi-particles and local
magnetic excitations in this compounds have been confirmed by recent
neutron scattering experiments \cite{Hiess04}. Theoretical studies
aiming at an explanation of the complex low-temperature structures
lay emphasis on the partitioning of the electronic density into localized
and delocalized parts \cite{Petit03,Wills04}. Considering the success
of the dual model in describing the low-energy behavior of actinide
heavy fermion compounds we have to explain how the splitting of the
$5f$ shells arises and propose means of identifying the underlying
microscopic mechanism.

The dual model as sketched above is an effective Hamiltonian for the
low-energy regime. In the spirit of Wilson's renormalization group
it should be obtained by integrating out processes at higher energies.
The central theoretical task is to identify the microscopic mechanisms
which cause the $5f$ band widths to renormalize to almost zero for
certain orbital symmetries while staying finite for others. In the
present paper we concentrate on the role of intra-atomic correlations,
the motivation being as follows: Recent model studies for the ground
state properties of small clusters show that intra-atomic correlations
as described by Hund's rules may strongly enhance anisotropies in
the kinetic energy and thus lead to an orbital-selective Mott transition
in $5f$-systems \cite{Zwicknagl03,efremov04}. This scenario closely
parallels the one postulated recently to explain the behavior of transition
metal oxides \cite{Anisimov02,Koga04,Liebsch05,Laad05a}. Here we
discuss how intra-atomic correlations affect the single-particle spectral
functions. The conjecture is that the above mentioned enhancement
of anisotropies in the kinetic energy can explain the dual nature
of the $5f$ electrons. In a real material this enhancement would
imply anisotropies far beyond these predicted by a standard electronic
structure calculation.

The concept of correlation-driven partial localization in U compounds
has been challenged by various authors (see e. g. \cite{Opahle04}).
The conclusions are drawn from the fact that conventional band structure
calculations within the Local Density Approximation (LDA) which treat
all $5f$-states as itinerant can reproduce ground state properties
like Fermi surface topologies, densities. The calculation of ground
state properties, however, cannot provide conclusive evidence for
the delocalized or localized character of the $5f$-states in actinides.
First, the presence of localized states can be simulated in standard
band calculations by filled bands lying (sufficiently far) below the
Fermi level. Second, the Fermi surface is mainly determined by the
number of particles in partially filled bands and the geometry of
the lattice which affects the dispersion of the conduction electrons.
A change in the number of band electrons by an even amount does not
necessarily affect the Fermi surface since a change by an even number
may correspond to adding or removing a filled band. As such, the Fermi
surface is not a sensitive test of the microscopic character of the
states involved. 

Following up this track of thought we would like to present here a
qualitative discussion of strongly correlated $5f$ systems emphasizing
properties of the spectral functions which would allow an experimental
discrimination between correlation-driven partial localization and
a usual filled-band scenario. We lay emphasis on the role of orbital
degeneracies and on intra-atomic correlations which compete with the
(anisotropic) kinetic energy. The features chosen for discussion are
the distribution of spectral weight, the formation of coherent bands
and the opening of an excitation gap.

We begin the discussion in Section \ref{sec:ModelHamiltonian} with
a brief exposition of the model Hamiltonian and the techniques emphasizing
the general structures involved. The numerical procedure and the parameters
adopted in the actual calculation are the presented in Section \ref{sec:ComputationalDetails}.
In the subsequent Sections \ref{sec:SpectralFunctionsClusters} and
\ref{sec:CPT}, we delineate details of the models and the results
obtained. The conclusions are summarized in Section \ref{sec:Summary}.

\section{Model Hamiltonian and Computational Method}

\label{sec:ModelHamiltonian}

For the adoption of a model describing the $5f$ electrons in actinide
based heavy fermion systems we rely on experience with ab initio electronic
structure calculations. In these materials, the direct overlap of
the $5f$ wave functions at neighboring sites is anticipated to be
rather small due to the large U-U distance. Detailed studies \cite{Albers86}
suggest that the $5f$ states acquire their dispersion by hybridization
with high-lying empty conduction states. We model these processes
by introducing weak effective transfer integrals. The interplay between
intra-atomic Coulomb interaction and anisotropic kinetic energy in
$5f$-systems is described by the simple model Hamiltonian \cite{efremov04}

\begin{equation}
H=H_{\mbox{\tiny\textrm{band}}}+H_{\mbox{\tiny Coul}}\;.\label{eq:hamilton}\end{equation}
 The local Coulomb repulsion

\begin{eqnarray}
H_{\mbox{\tiny Coul}} & = & \frac{1}{2}\sum_{a}\sum_{j_{z_{1}},\dots,j_{z_{4}}}U_{j_{z_{1}}j_{z_{2}}j_{z_{3}}j_{z_{4}}}\nonumber \\
 &  & \times c_{j_{z_{1}}}^{\dag}(a)c_{j_{z_{2}}}^{\dag}(a)c_{j_{z_{3}}}(a)c_{j_{z_{4}}}(a).\label{eq:hcoul}\end{eqnarray}
 is written in terms of the usual fermionic operators $c_{j_{z}}^{\dagger}(a)$
($c_{j_{z}}(a)$) which create (annihilate) an electron at site $a$
in the $5f$-state with total angular momentum $j$ and $z$-projection
$j_{z}$. Considering the fact that the spin-orbit splitting is large
we neglect contributions from the excited spin-orbit multiplet $j=7/2$
and adopt the $j-j-$coupling scheme. The Coulomb matrix elements
$U_{j_{z_{1}}j_{z_{2}}j_{z_{3}}j_{z_{4}}}$ for $j_{zi}=-5/2,\ldots,5/2,$

\begin{eqnarray}
\lefteqn{U_{j_{z_{1}}j_{z_{2}}j_{z_{3}}j_{z_{4}}}=\delta_{j_{z_{1}}+j_{z_{2}},j_{z_{3}}+j_{z_{4}}}}\nonumber \\
 &  & \times\sum_{J}U_{J}\, C_{5/2,j_{z1};5/2,j_{z2}}^{JJ_{z}}C_{5/2,j_{z3};5/2,j_{z4}}^{JJ_{z}}.\end{eqnarray}
 are given in terms of the usual Clebsch-Gordan coefficients $C_{\ldots}^{\ldots}$
and the Coulomb parameters $U_{J}$.

The kinetic energy operator in the band term $H_{\mbox{\tiny band}}$
describes the hopping between all pairs neighboring sites $\langle ab\rangle$

\begin{eqnarray}
H_{\mbox{\tiny band}} & = & -\sum_{\langle ab\rangle,j_{z}}t_{j_{z}}\left(c_{j_{z}}^{\dag}(a)c_{j_{z}}(b)+\mbox{h.c.}\right)\nonumber \\
 &  & +\sum_{a,j_{z}}\epsilon_{f}c_{j_{z}}^{\dag}(a)c_{j_{z}}(a)\quad.\label{eq:Hband}\end{eqnarray}
 We assume the transfer integrals $t_{j_{z}}$ to be diagonal in the
orbital index $j_{z}$. This simplification is justified for a 1-dimensional
system since the cylindrical symmetry allows one to select a common
quantization axis for all sites. We would like to stress that although
the hopping term is assumed to be diagonal in the orbital indicies
the corresponing channels are coupled through the Coulomb term. Finally,
we account for the orbital energy $\epsilon_{f}$ which determines
the $f-$valence of the ground state.

The single-particle spectral function \begin{eqnarray}
\lefteqn{A_{j_{z}}(k,\omega)=}\nonumber \\
 & = & \begin{cases}
\sum_{n}\left|\left\langle \Psi_{n}^{(N+1)}\left|c_{j_{z}}^{\dagger}(k)\right|\Psi_{0}^{(N)}\right\rangle \right|^{2}\delta(\omega-\omega_{n0}^{(+)}) & ;\textrm{ }\omega>0\\
\sum_{n}\left|\left\langle \Psi_{n}^{(N-1)}\left|c_{j_{z}}(k)\right|\Psi_{0}^{(N)}\right\rangle \right|^{2}\delta(\omega+\omega_{n0}^{(-)}) & ;\textrm{ }\omega<0\end{cases}\nonumber \\
\label{eq:DefinitionSpectralFunction}\end{eqnarray}
 with \begin{equation}
c_{j_{z}}(k)=\frac{1}{\sqrt{L}}\sum_{a}e^{ika}c_{j_{z}}(a)\;;\;\omega_{n0}^{(\pm)}=E_{n}^{(N\pm1)}-E_{0}^{(N)}\end{equation}
yields the probability for adding ($\omega>0$) or removing ($\omega<0$)
an electron with energy $\omega$ in a state characterized by momentum
$k$ and angular momentum projection $j_{z}$ to the $N$-particle
ground state $\left|\Psi_{0}^{(N)}\right\rangle $ with energy $E_{0}^{(N)}$.
The states with $N\pm1$ and their energies are denoted by $\left|\Psi_{n}^{(N\pm1)}\right\rangle $
and $E_{n}^{(N\pm1)}$, respectively, and $L$ is the number of sites.

The importance of the spectral functions stems, first, from the fact
that they can be observed in photoemission and inverse photoemission
experiments and second, that the orbital-projected expectation value
of the kinetic energy per site\begin{equation}
T_{j_{z}}=\int_{-\infty}^{+\infty}d\omega\, f(\omega)\,\frac{1}{L}\sum_{k}\epsilon_{j_{z}}(k)A_{j_{z}}(k,\omega)\end{equation}
 as well as the orbital-projected momentum distribution function\begin{equation}
n_{j_{z}}(k)=\int_{-\infty}^{+\infty}d\omega\, f(\omega)A_{j_{z}}(k,\omega)\label{eq:MomentumDistributionFunction}\end{equation}
 and the orbital-projected Density of States (DOS) \begin{equation}
D_{j_{z}}(\omega)=\frac{1}{L}\sum_{k}A_{j_{z}}(k,\omega)\end{equation}
 can be expressed through them. The dispersion of the band energies
$\epsilon_{j_{z}}(k)$ is given by the kinetic energy Eq. (\ref{eq:Hband}),
$L$ denotes the number of lattice (cluster) sites while $f$ is the
usual Fermi function. 

\begin{figure}[tb]
\includegraphics[%
  width=0.80\columnwidth]{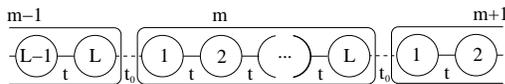}

\caption{Representation of a one-dimensional chain as a super-lattice of identical
$L$-site cluster. The orbital-dependent transfer integrals between
neighboring sites within a cluster and between adjacent clusters are
$t=(t_{j_{z}=-5/2},\dots,t_{j_{z}=5/2})$ and $t_{0}=(t_{0j_{z}=-5/2},\dots,t_{0j_{z}=5/2})$,
respectively. \label{fig:CPT} }
\end{figure}

We determine the spectral function $A_{j_{z}}(k,\omega)$ from the
single particle Green's function $G_{j_{z}}(k,\omega)$,

\begin{equation}
A_{j_{z}}(k,\omega)=-\frac{1}{\pi}\lim_{\eta\rightarrow0^{+}}\textrm{ Im }G_{j_{z}}(k,\omega+i\eta)\end{equation}
 which, in turn, are calculated applying Cluster Perturbation Theory
(CPT) \cite{sen02,sen99,sennk,pai99,gros04}. The method which has
been successfully applied to a wide variety of many-particle models
proceeds by dividing the infinite periodic lattice into identical
finite clusters as illustrated in Figure \ref{fig:CPT}. %

The Hamiltonian (\ref{eq:hamilton}) takes the form

\begin{equation}
H=\sum_{m}H_{m}^{0}+\sum_{m,n,a,b,j_{z}}V_{abj_{z}}^{mn}c_{maj_{z}}^{\dag}c_{nbj_{z}},\label{eq:SuperlatticeHamiltonian}\end{equation}
 where $H_{m}^{0}$ refers to cluster $m$ while $V_{abj_{z}}^{mn}$
describes the hopping between sites $a$ and $b$ on the adjacent
clusters $m$ and $n$. The Green's functions of the finite clusters\begin{equation}
\left(\hat{G}_{j_{z}}^{(C)}(z)\right)_{ab}=G_{abj_{z}}^{(C)}(z)\;;\; a,b=1,\ldots,L\label{eq:ClusterGreensFunction}\end{equation}
 are calculated exactly thereby fully accounting for the complex dynamics
of the strongly correlated system. Here we use the locator representation
where the indices refer to the sites in the cluster. In a second step,
the inter-cluster hopping $V_{abj_{z}}^{mn}$, is treated as weak
perturbation which yields the Green's function of the super-lattice
\cite{sen99,sen02}

\begin{equation}
G_{abj_{z}}(Q,z)=\left(\frac{\hat{G}_{j_{z}}^{(C)}(z)}{1-\hat{V}_{j_{z}}(Q)\hat{G}_{j_{z}}^{(C)}(z)}\right)_{ab}\quad.\label{eq:GreenCPT}\end{equation}
 The variation with the wave vector $Q$ which is taken from the first
Brillouin zone of the super-lattice is introduced by the inter-cluster
hopping matrix $\hat{V}_{j_{z}}(Q)$. Its matrix elements are obtained
from $V_{abj_{z}}^{mn}$ by performing a Fourier transformation with
respect to the cluster indices $m$ and $n$. Following \cite{sen02},
the translational invariance of the original lattice is explicitly
restored yielding the Green's function

\begin{equation}
G_{j_{z}}(k,z)=\frac{1}{L}\sum_{a,b}e^{-ik(a-b)}G_{abj_{z}}(Lk,z)\quad.\label{eq:CPTFinalFourier}\end{equation}
Note that the CPT is exact in three limiting cases: vanishing hopping
terms, in an uncorrelated system and for infinite cluster size. The
approximation is controlled most effectively by considering different
cluster sizes $L$ 

Before we turn to a detailed description of the calculations we would
like to mention the following caveat. The formulation of CPT as summarized
above implicitly relies on the assumption that the ground states of
the interacting electron system are non-degenerate. This proposition,
however, may not be satisfied for the building blocks of the super-lattice.
The problem arises trivially for clusters containing an odd number
of electrons where all states are at least two-fold degenerate. Of
course one could try to circumvent the problem by choosing appropriate
cluster sizes or by varying the band filling, i. e., the number of
particles. This strategy which may work for simple systems does not
provide a solution in orbitally degenerate systems. Due to the local
orbital degrees of freedom the complexity increases rapidly with the
number of sites or particles imposing serious restrictions on models.
As we attempt at a description of homogeneous phases we separately
calculate the Green's functions for the states of the ground state
manifold and subsequently average the spectral function.

\section{Computational Details}

\label{sec:ComputationalDetails}

We begin by specifying the model parameters entering the Hamiltonian
Eq. (\ref{eq:hamilton}). The high-energy scale is set by the Coulomb
repulsion between two $5f$ electrons at the same site. The parameters
$U_{J}$ are chosen according to the following considerations. First,
it is well known, that the isotropically averaged Coulomb repulsion
in a metal is strongly reduced as compared to its value in an atom.
The reduction is a direct consequence of screening by the itinerant
conduction electrons which, in turn, implies that the actual value
depends upon the chemical environment of the correlated ion under
consideration. We do not attempt at an ab initio calculation this
quantity but rather leave it as a parameter, $U_{at}$, whose value
can be estimated from the positions of the valence peaks in photoemission
and inverse photoemission. The intra-atomic correlation which are
the focus of interest in the present paper involve the anisotropic
parts of the Coulomb interaction. The latter give rise to the multiplet
structure. It is important to note that the corresponding interactions
are (usually) not screened and hence retain their atomic values. As
we expect the anisotropic Coulomb parameters to be rather robust with
respect to changes in the chemical composition or to pressure we use
fixed values for the differences 

\begin{eqnarray}
\Delta U_{4} & = & U_{J=4}-U_{J=0}=-3.79\textrm{eV}\nonumber \\
\Delta U_{2} & = & U_{J=2}-U_{J=0}=-2.72\textrm{eV}\quad.\label{eq:HundEnergies}\end{eqnarray}
 which were determined from an ab initio calculation for UPt$_{3}$
\cite{Zwicknagl02}. The value for the orbital energy $\epsilon_{f}$
in Eq. (\ref{eq:Hband}) is fixed by the requirement that low energy
states are formed by linear combinations of the ionic $f^{2}$- and
$f^{3}$-configurations which are energetically (almost) degenerate.
The intermediate valent ground state with $n_{f}\simeq2.5$ allows
for low-energy valence transitions and the formation of bands also
in the strong-coupling limit.

Finally, the values of the transfer integrals t$_{j_{z}}$ are varied
in the strong-coupling regime. As in Ref. \cite{efremov04} we choose
t$_{1/2}$=t$_{5/2}$ in the figures.

We next turn to the calculation of the cluster Green's functions $\hat{G}_{j_{z}}^{(C)}(z)$
which account for the local correlated dynamics. We should like to
mention that the use of the open boundary conditions required by the
CPT method considerably increases the numerical effort over that encountered
in finite clusters with periodic boundary conditions. A further difficulty
arises from the fact that the orbital degrees of freedom may lead
to a rather high density of low-energy states. Keeping in mind these
complications the Green's functions for clusters with two and three
correlated sites are calculated using Lanczos continued-fraction technique
\cite{Dagotto94}. To avoid convergence problems due to the near degeneracy
of the lowest excited states in the $5f$ systems we use the \emph{JDQR}
algorithm \cite{slei99} instead of the usual Lanczos method to obtain
the ground state. 

For purposes of comparison we present alongside the CPT spectra the
test cases of finite clusters using periodic boundary conditions.
In this case, the calculation of the cluster Green's function $\hat{G}_{j_{z}}^{(C)}(z)$
simplifies to\begin{equation}
G_{abj_{z}}^{(C)}(z)=\frac{1}{L}\sum_{k_{\ell=1}}^{k_{L}}e^{ik_{\ell}(a-b)}G_{j_{z}}^{(C)}(k_{\ell},z)\end{equation}
where \cite{FuldeBook}

\begin{eqnarray}
G_{j_{z}}^{(C)}(k,z) & = & G_{j_{z},e}^{(C)}(k,z)+G_{j_{z},h}^{(C)}(k,z)\nonumber \\
G_{j_{z},e}^{(C)}(k,z) & = & \langle\Psi_{0}|c_{j_{z}}(k)\,\frac{1}{z-H+E_{0}}\, c_{j_{z}}^{\dagger}(k)|\Psi_{0}\rangle\nonumber \\
G_{j_{z},h}^{(C)}(k,z) & = & \langle\Psi_{0}|c_{j_{z}}^{\dagger}(k)\,\frac{1}{z+H-E_{0}}\, c_{j_{z}}(k)|\Psi_{0}\rangle\quad.\label{eq:ClusterGreen}\end{eqnarray}
The discrete value $k_{\ell}=\ell\frac{2\pi}{L}\;;\ell=0,\ldots,L-1$
label the single-particle eigenstates of the $L$-site cluster. The
Green's functions are evaluated for complex frequencies $z=\omega+i\eta$
with $\eta>0$. The imaginary part leads to additional Lorentzian
broadening of the spectral functions and may affect the behavior of
the integrals like the expectation value of the kinetic energy or
the momentum distribution function. All these quantities are increasingly
difficult to compute as $\eta$ decreases. For the actual calculations
we choose values for $\eta$ of the order of $10^{-2}$ of the relevant
hopping integrals. For the plots a rather large value of $\eta=0.03$
is chosen to allow for a good visualization. 

The absolute values of $U_{J}$ are very large compared to the hopping
integral and thus we shall limit ourselves to the subspace of $f^{2}$
and $f^{3}$ configurations for the ground state which has shown to
be a good approximation. This leads to a considerable reduction of
Hilbert space dimension and less computational costs. The actual size
of the Hilbert spaces in the reduced systems for the two site cluster
filled with five electrons are $600$ states and $220$ respectively
$400$ for four and five electrons. For the clusters with three sites
one obtains $18\ 000$ dimensions for a filling with eight electrons
and $8\ 000$ respectively $13\ 500$ for seven and nine electrons.
The matrix representations of the Hamiltonian are then stored as sparse
matrices.

\section{Spectral Functions of Finite Clusters}

\label{sec:SpectralFunctionsClusters}

\subsection{Atomic limit}

We begin by discussing the spectral function of our model in the atomic
limit neglecting hopping between the sites. The results provide a
first qualitative insight into the behavior in the strong-coupling
limit. Due to the rotational invariance of the Coulomb interaction
Eq. (\ref{eq:hcoul}) the spectral functions $A_{j_{z}}(\omega)$
will not depend on the magnetic quantum number $j_{z}$. The lowest-energy
states of an intermediate-valent system in the zero-band width limit
are given by products of atomic $f^{2}$- and $f^{3}$-configurations
whose energies are assumed to be (almost) degenerate. The corresponding
spectral functions are obtained in close analogy to the classical
work by Hubbard \cite{HubbardII}. The zero-configuration width approximation
which neglects intra-atomic correlations leads to a characteristic
three-peak structure. The valence transitions $f^{2}\rightarrow f^{1}$
and $f^{3}\rightarrow f^{4}$ occur at large energies and, concomitantly,
do not affect the low-temperature behavior. The latter is determined
by the low-energy peak resulting from the transitions $f^{2}\leftrightarrow f^{3}$
within the $f^{2}$- and $f^{3}$-configurations. This peak is a direct
consequence of the intermediate-valent ground state. The strong correlations
present imply a substantial transfer of spectral weight from the low-energy
part to the high-energy regime. The weights of the peaks can be estimated
from combinatorial considerations. The weight $Z(f^{2}\rightarrow f^{1})$
of the transition $f^{2}\rightarrow f^{1}$ equals the probability
that a state with a given $j_{z}$ is occupied in that $f^{2}$ contribution
of the mixed-valent ground-state. Following these lines one finds:

\begin{eqnarray}
Z(f^{2}\rightarrow f^{1})=\frac{1}{6} & \quad & Z(f^{2}\rightarrow f^{3})=\frac{1}{3}\nonumber \\
Z(f^{3}\rightarrow f^{2})=\frac{1}{4} & \quad & Z(f^{3}\rightarrow f^{4})=\frac{1}{4}\label{eq:WeightsForValenceTransitions}\end{eqnarray}

The central focus of the present paper is the evolution of the low-energy
peak whose spectral weight sums up to $7/12$. Intra-atomic correlations
which are usually described by Hund's rules further reduce the spectral
weight of the low-energy excitations involving valence transitions
$\left|f^{3},J=9/2\right\rangle \rightleftharpoons\left|f^{2},J=4\right\rangle $
between the ground state multiplets. The corresponding spectral weights
can be expressed in a straightforward way in terms of the usual Clebsch-Gordan
coefficients and the reduced matrix elements. In addition to the central
peak one finds transitions to the excited multiplets $\left|f^{3},J=9/2\right\rangle \rightarrow\left|f^{2},J=2\right\rangle $
and $\left|f^{2},J=4\right\rangle \rightarrow\left|f^{3},J=3/2\right\rangle \;,\;\left|f^{3},J=5/2\right\rangle $
occurring at the corresponding multiplet excitations. The excitation
energies as well as the corresponding spectral weights are listed
in Table \ref{cap:Positions-and-spectral}. We should like to mention
that there is no transition into the excited multiplet $\left|f^{2},J=0\right\rangle $.

\begin{table}[tb]
\begin{tabular}{ccccccc}
\hline 
&
\multicolumn{2}{c}{PES}&
&
\multicolumn{3}{c}{BIS}\tabularnewline
\hline
Position / eV&
 -1.06&
 0.00&
&
 0.00&
 1.60&
 2.73\tabularnewline
 Spectral weight&
 0.054&
 0.196&
&
 0.218&
 0.032&
 0.083 \tabularnewline
\hline
\end{tabular}

\caption{Positions and spectral weight of the atomic transitions in uranium
from numerical calculations. For $\omega\leq0$ eV the contributions
are from the hole propagation and for $\omega\geq0$ eV from the particle
propagation.\label{cap:Positions-and-spectral}}
\end{table}

\subsection{Results for finite clusters}

\begin{figure}[b]
\includegraphics[%
  width=0.80\columnwidth]{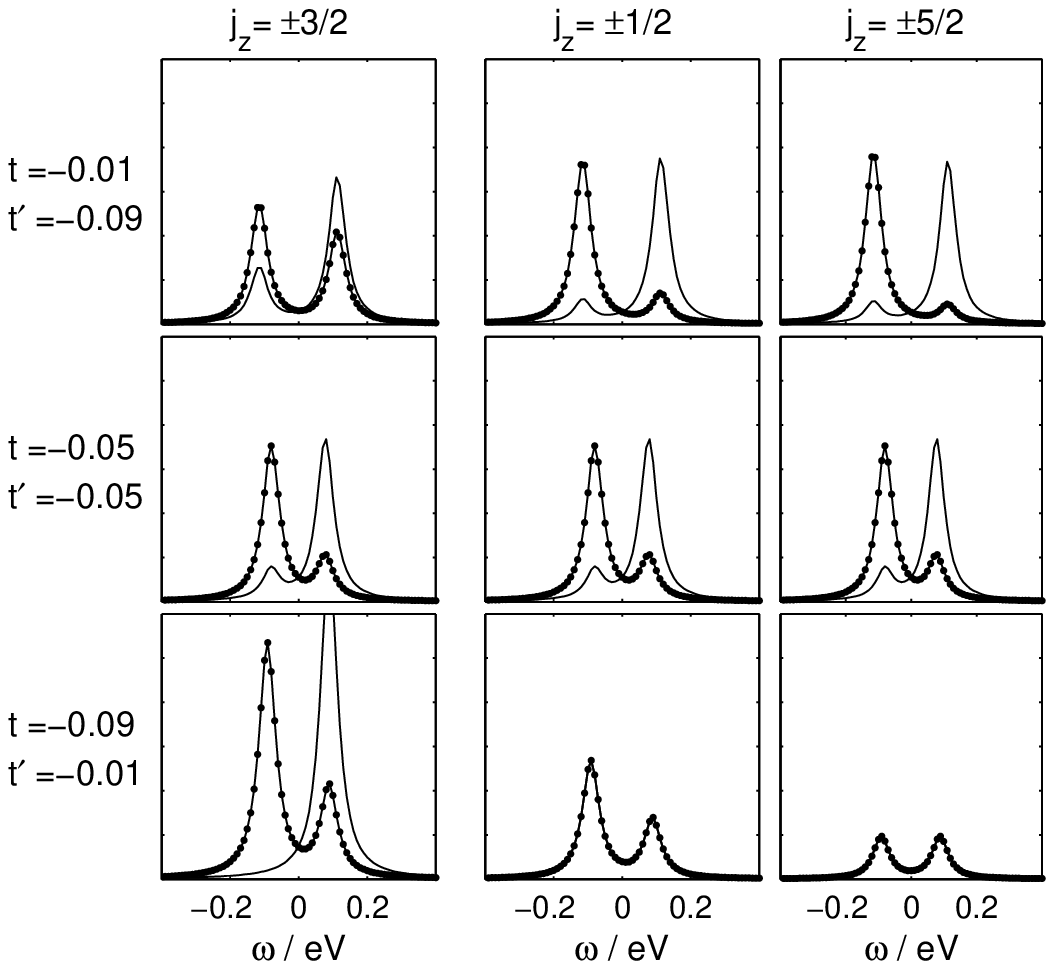}

\caption{Variation with wave number of the orbital-projected spectral functions
calculated for a two-site cluster with five electrons and the hopping
parameters $t_{3/2}=t$ and $t_{1/2}=t_{5/2}=t'$. The full line and
the dotted line refer to $k=0$ and $k=\pi$, respectively. The spectra
are obtained from averaging the Green's functions over the degenerate
ground state manifold. Spectral weight is transferred to local excitations
(valence transitions and transitions into excited atomic multiplets)
which are not displayed here. The Lorentzian broadening is $\eta=0.03$\label{cap:LowEnergySpectraTwoSiteCluster}.}
\end{figure}
\begin{figure}[bh]
\includegraphics[%
  width=0.80\columnwidth]{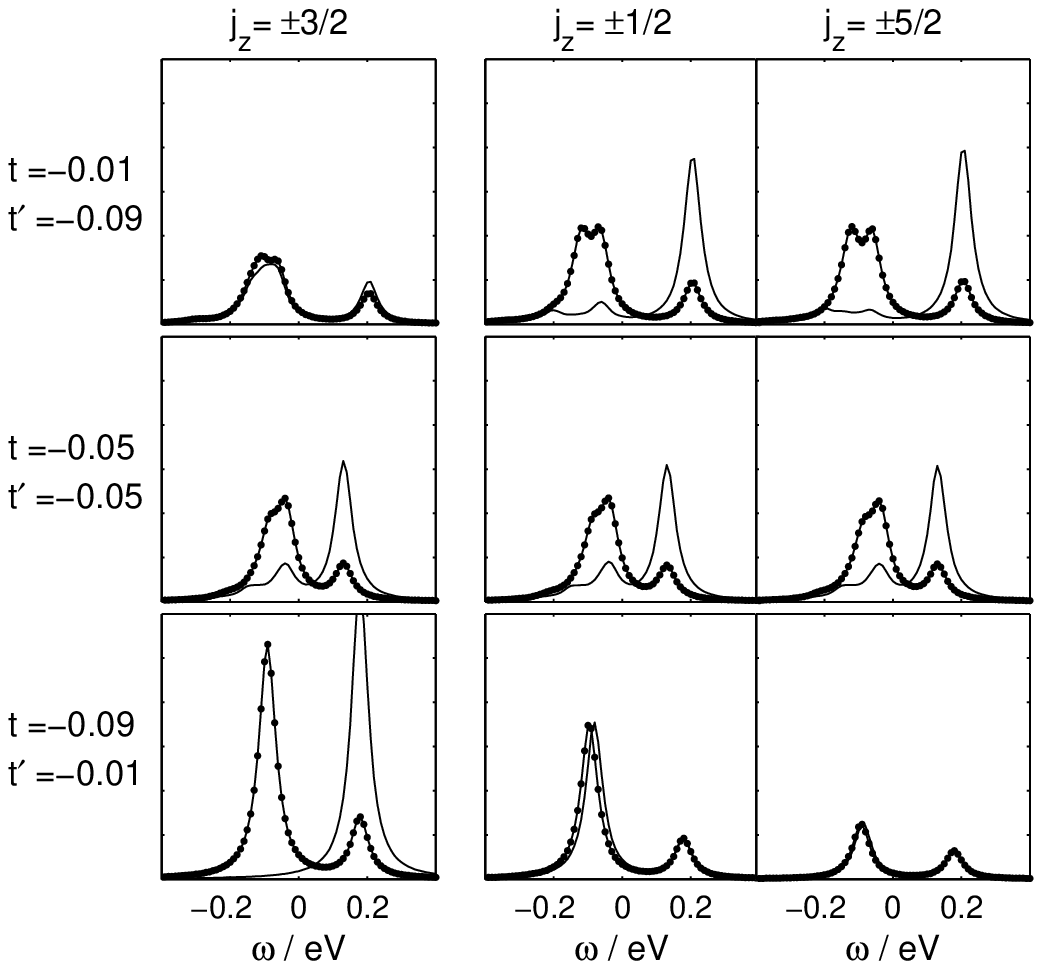}

\caption{Variation with wave number of the orbital-projected spectral functions
calculated for a three-site cluster with eight electrons and the hopping
parameters $t_{3/2}=t$ and $t_{1/2}=t_{5/2}=t'$. The full line and
the dotted line refer to $k=0$ and $k=\pm\frac{2\pi}{3}$, respectively.
The spectra are obtained from averaging the Green's functions over
the degenerate ground state manifold. Spectral weight is transferred
to local excitations (valence transitions and transitions into excited
atomic multiplets) which are not displayed here. The Lorentzian broadening
is $\eta=0.03$\label{cap:LowEnergySpectraThreeSiteCluster}.}
\end{figure}
\begin{figure}[t]
\subfigure[]{\includegraphics[%
  width=0.48\columnwidth]{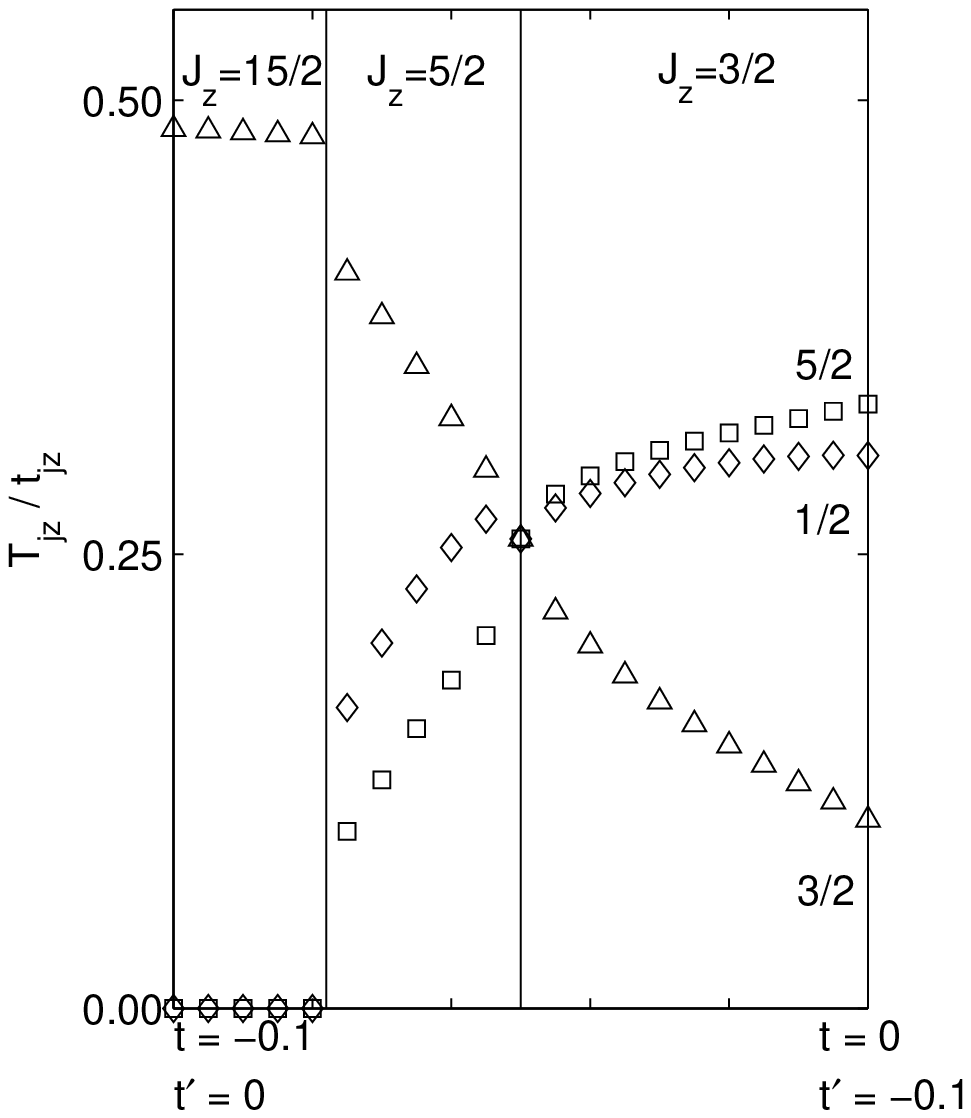}} \subfigure[]{\includegraphics[%
  width=0.48\columnwidth]{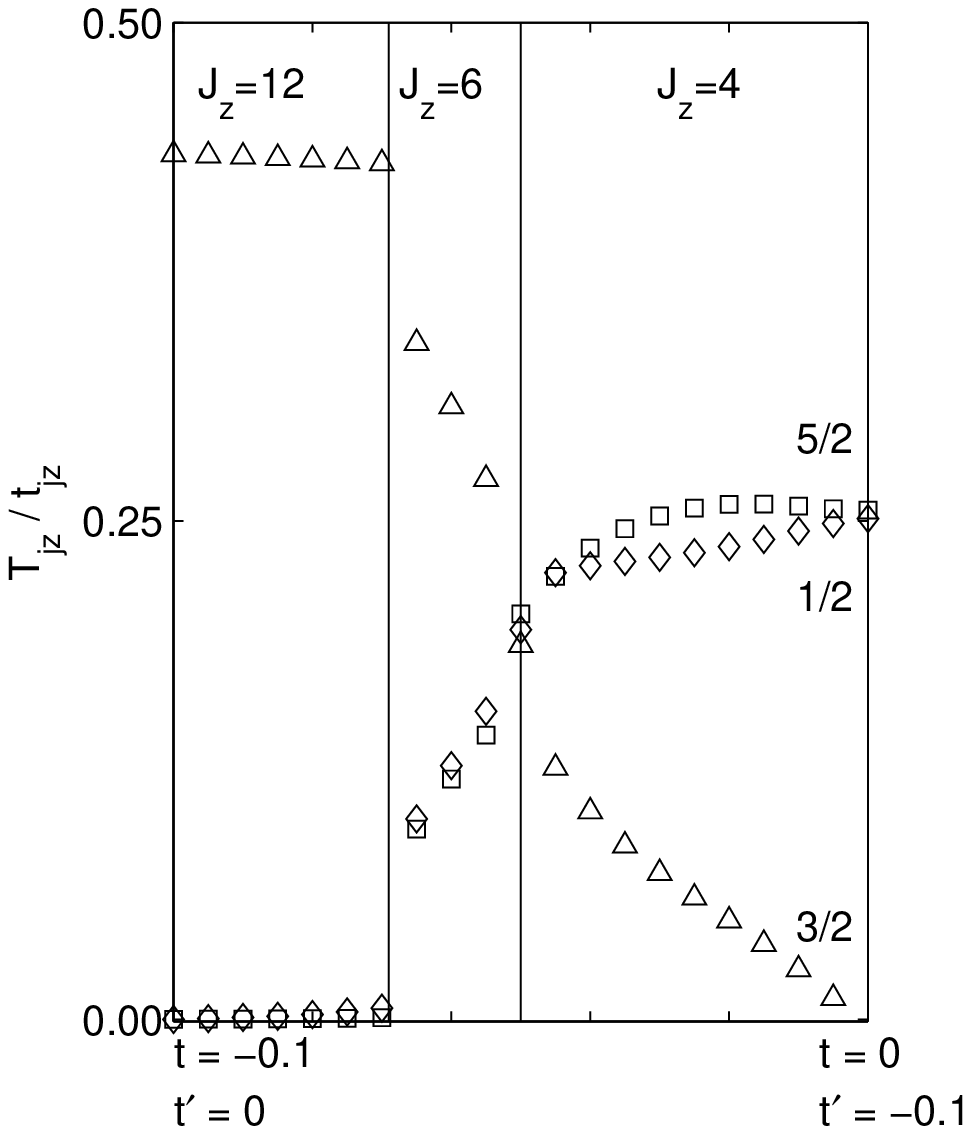}}

\caption{Ratio $T_{j_{z}}/t_{j_{jz}}$ of the two site cluster (left) and
three site cluster (right) with $t_{j_{z}}<0$ along the line $t_{3/2}=t$
and $t_{1/2}=t_{5/2}=t'$ connecting the values below the figures.
\label{cap:projected} }
\end{figure}

We next turn to the modifications introduced by the hopping between
different sites which acts as a weak perturbation in the strong-coupling
limit. Its principal effect is to remove the rotational symmetry of
the isolated atom and, consequently, remove the high degeneracy of
the ground state. As the total angular momentum ceases to be a good
quantum number the ground state may contain admixtures from excited
$J$-multiplets. The model Hamiltonian as specified by Eqs. (\ref{eq:hamilton})-(\ref{eq:Hband}),
however, conserves the z-component of the angular momentum allowing
us to classify the eigenstates with respect to $J_{z}$. The magnetic
character of the ground states depends upon the transfer integrals
as shown in Ref. \cite{efremov04}. These modifications affect the
spectral functions $A_{j_{z}}(k_{\ell},\omega)$ of an L-site cluster
where the discrete set of quantum numbers $k_{\ell}=0,\dots,(L-1)\frac{2\pi}{L}$
labels the single-particle eigenstates. First, the positions of the
peaks corresponding to the transitions between different valence states
may acquire dispersion. Second, spectral weight may be transferred
from the low-energy regime to high-energy satellites. Third, transitions
to excited multiplets not present in the atomic limit may appear and,
finally, the overall spectral weight may be redistributed among the
different $j_{z}$-channels. These issues will be addressed in the
following section. 

A first qualitative understanding can be gained by considering the
spectral functions of a two-site cluster where simple approximate
expressions are found for the ground states in limiting cases. Throughout
the discussion we restrict ourselves to the strong-coupling regime
$\left|t_{j_{z}}\right|\ll\left|\Delta U_{J}\right|$ where the ground
state is mainly built from products of local $f^{2}$ configurations
$\left|f^{2};4J_{z}\right\rangle $ and local $f^{3}$ configurations
$\left|f^{3};9/2J'_{z}\right\rangle $. Finite hopping between the
two sites $a$ and $b$ splits the manifold in first order perturbation
theory. We consider the case of strongly anisotropic hopping $\left|t_{3/2}\right|=\left|t\right|\gg\left|t'\right|=\left|t_{5/2}\right|=\left|t_{1/2}\right|$
and $\left|t_{3/2}\right|=\left|t\right|\ll\left|t'\right|=\left|t_{5/2}\right|=\left|t_{1/2}\right|$
as well as isotropic hopping $t_{3/2}=t=t'=t_{5/2}=t_{1/2}$. The
variation with $t$, $t'$ of the spectral functions is displayed
in Figure \ref{cap:LowEnergySpectraTwoSiteCluster}. 

Let us first consider the behavior along the isotropic line $t_{3/2}=t=t'=t_{5/2}=t_{1/2}$.
Due to the rotationally invariance, the magnitude $\mathcal{J}^{2}$
of the total angular momentum $\mathcal{J}=J(a)+J(a)$ provides a
good quantum number. It has been shown previously \cite{efremov04}
that the ground state is six-fold degenerate. It is a linear combinations
of the $\mathcal{J}^{2}$ eigenstates $\left|(\frac{9}{2}\,4)\mathcal{JJ}_{z}\right\rangle $
and $\left|(4\,9)\mathcal{JJ}_{2z}\right\rangle $ for $\mathcal{J}=5/2$
which can be expressed in terms of the above-mentioned product states
$\left|f^{3};\frac{9}{2}\mathcal{J}_{z}-J_{z}\right\rangle \left|f^{2};4J_{z}\right\rangle $
and $\left|f^{2};4J_{z}\right\rangle \left|f^{3};\frac{9}{2}\mathcal{J}_{z}-J_{z}\right\rangle $.
The gain in kinetic energy\begin{eqnarray}
\left|t\right| & \left|\left\langle \left(\frac{9}{2}\,4\right)\mathcal{JJ}_{z}\left|\sum_{j_{z}}c_{j_{z}}^{\dagger}(a)c_{j_{z}}(b)\right|\left(4\,\frac{9}{2}\right)\mathcal{JJ}_{z}\right\rangle \right|\nonumber \\
 & =\left|t\right|\frac{33}{14}\left|\tau(\mathcal{J})\right|\label{eq:KinEnergyIsotropicCase}\end{eqnarray}
given in terms of the reduced matrix element $\left(\frac{9}{2}\left\Vert c_{j_{z}}^{\dagger}\right\Vert 4\right)=\sqrt{\frac{33}{14}}$
and the $6j$-symbol\begin{equation}
\tau(\mathcal{J})=(-1)^{\mathcal{J}}10\left(\begin{array}{ccc}
4 & \frac{5}{2} & \frac{9}{2}\\
4 & \mathcal{J} & \frac{9}{2}\end{array}\right)\end{equation}
 for $\mathcal{J}=5/2$ results from the formation of symmetric (antisymmetric)
combinations the above-mentioned $\mathcal{J}^{2}$ eigenstates. The
isotropy of the ground state is reflected in the isotropic spectral
functions which exhibit peaks at the energies \begin{equation}
\pm\frac{33}{14}\,\frac{907}{1386}\, t=\pm0.077\textrm{ eV}\quad.\end{equation}
 Due to the strong correlations we find two peaks for both removing
(adding) electrons in symmetric states $c_{j_{z}}^{\dagger}(k=0)$
and antisymmetric states $c_{j_{z}}^{\dagger}(k=\pi)$. 

In the strongly anisotropic case $\left|t_{3/2}\right|=\left|t\right|\gg\left|t'\right|=\left|t_{5/2}\right|=\left|t_{1/2}\right|$,
the wave function of the Kramers' degenerate ground state with $\mathcal{J}_{z}=\pm15/2$
has the simple form (for $\mathcal{J}_{z}=+15/2$)\begin{eqnarray}
\left|\Psi\right\rangle _{15/2} & = & \frac{1}{\sqrt{2}}\left(c_{3/2}^{\dagger}(a)+\frac{t}{|t|}\, c_{3/2}^{\dagger}(b)\right)\times\nonumber \\
 & = & c_{5/2}^{\dagger}(a)c_{1/2}^{\dagger}(a)c_{5/2}^{\dagger}(b)c_{1/2}^{\dagger}(b)\label{eq:Psi015By2}\end{eqnarray}
describing an independent $j_{z}=3/2$-electron in a magnetically
polarized background. This state is separately an eigenstate of the
kinetic energy as well as the Coulomb energy. Removing an electron
in the single-particle state $\frac{1}{\sqrt{2}}\left(c_{3/2}^{\dagger}(a)+\frac{t}{|t|}\, c_{3/2}^{\dagger}(b)\right)$
as well as adding one in its orthogonal counterpart leads to final
states $J(a)=4=J(b)$ and $J(a)=9/2=J(b)$, respectively, where both
sites are in the ground state multiplets of the local $f^{2}$ and
$f^{3}$ configurations. As a consequence, we find two sharp peaks
with full spectral weight at the energies $\omega=\pm|t|$ which evolve
into a quasiparticle band in extended systems. The additional spectral
weight appearing in Figure \ref{cap:LowEnergySpectraTwoSiteCluster}
at positive energies $\omega=+|t|$ accounts for the probability of
adding an electron in the minority channel $j_{z}=-3/2$. This generates
final states which contain contributions from excited states like
unfavorable local $f^{4}$-configurations and excited multiplets in
the local $f^{3}$-subspaces. The overlap with states from the local
ground state manifold gives rise to a peak at $\omega=+|t|$ whose
reduced height, however, highlights the transfer of significant spectral
weight to the high-energy regime. Finally, the absence of dispersion
in the $j_{z}=1/2$ and $j_{z}=5/2$ spectral functions as well as
the reduced spectral weight in the low-energy regime reveal the localized
character of the corresponding orbitals in the ground state. The evolution
of the spectra and in particular the reduction of spectral weight
in the low-energy regime is reminiscent of the gap formation at the
Mott-Hubbard transition. The qualitative difference between the $5/2$-
and the $1/2$-channel as well as the asymmetry within the $1/2$-channel
reflect the intra-atomic (Hund's rule type) correlations.

The structure of the spectral functions in the complementary anisotropic
limit $\left|t_{5/2}\right|=\left|t_{1/2}\right|=\left|t'\right|\gg\left|t\right|=\left|t_{3/2}\right|$
can be explained by similar considerations. The Kramers' degenerate
ground state is built from configurations where all but the $j_{z}=-3/2$
($j_{z}=3/2$) orbitals are occupied. This implies $\mathcal{J}_{z}=\pm3/2$.
It is important to note that this implies antiferromagnetic correlations
between the sites. In this state, delocalization of the $j_{z}=3/2$
($j_{z}=-3/2$) is not principally excluded. It is, however, associated
with breaking Hund's rules. As a result we find dispersion in the
corresponiding channels. The spectral weight, however, is reduced
due to intra-atomic excitations. 

The characteristic structure of the spectra as summarized above is
present also in the three-site cluster as can be seen from Figure
\ref{cap:LowEnergySpectraThreeSiteCluster}. The coherence effects,
however, are more pronounced. For a discussion of the (approximate)
ground states in limiting cases we refer to \cite{efremov04}.

A striking feature is the splitting of the lower peaks for $j_{z}=\pm1/2$
and $j_{z}=\pm5/2$ at $t_{1/2}=t_{5/2}=-0.09$, $t_{3/2}=-0.01$.
The key to the explanation is provided by the observation that the
corresponding ground state is characterized by antiferromagnetic intersite
correlations, a property encountered already in the two-site cluster
discussed above. The implications for the three-site system, however,
are more subtle. For periodic boundary conditions the ground stae
is a complex superposition of several (almost) equivalent states reflecting
frustration. Removing an electron from the ground state leads to rather
complex final states which, inturn , give rise to the observed splitting. 

The most prominent feature of the spectral functions in the strongly
anisotropic limit is the partial localization which can be viewed
as an orbital-selective Mott transition. The intra-atomic correlations
strongly enhance the anisotropies of the kinetic energy as can be
seen from Figure \ref{cap:projected}. The calculated projected kinetic
energy $T_{j_{j}}/t_{j_{z}}$ (Figure \ref{cap:projected} ) agrees
well with results from ground state wavefunction \cite{efremov04}.
The small differences to the results from the ground state wave function
are due to Lorentzian broadening with $\eta=0.005$.

\section{Cluster Perturbation Theory}

\label{sec:CPT}

The essential benefit gained from CPT is the possibility to calculate
the spectral functions at arbitrary wave vectors and band filling.
In the present paper we explore the first aspect treating CPT as a
sophisticated interpolation scheme for the variation with $k$. The
averaged particle number per site, on the other hand, is kept fixed
at the value used in the underlying cluster calculation. Comparing
the spectra of the infinite chain to those of isoelectronic finite
clusters provides us with an estimate of finite size effects since
the building blocks of the chain are treated with different boundary
conditions.

\begin{figure}[t]
\includegraphics[%
  width=0.80\columnwidth]{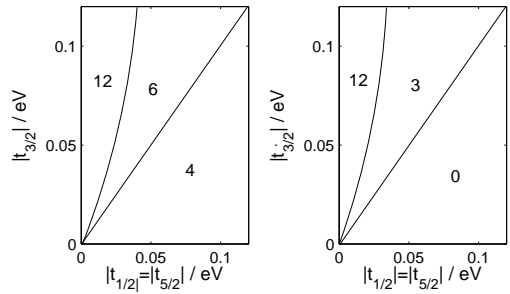}

\caption{Phase diagrams of the three-site cluster filled with 8-electrons
for negative hopping parameters $t_{j_{z}}$, derived for the total
magnetization $J_{z}$ for periodic boundary condition (left panel)
and open boundary conditions (right panel).\label{pd3p}}
\end{figure}

The qualitative behavior of the spectral functions reflects the detailed
structure of ground state which is determined by the competition between
the energy gain due to intra-atomic correlations and kinetic energy.
The latter, however, depends upon the boundary conditions imposed
upon the wave functions. The cluster calculations discussed in the
preceding section were performed adopting the usual periodic boundary
conditions. The CPT, however, requires cluster Green's function for
open boundary conditions. In the strong-coupling regime, the dependence
upon the transfer integrals of the ground state is only weakly affected
by the actual choice of the boundary conditions. In the strongly anisotropic
limit with one dominating transfer integral (a) the high-spin states
with ferromagnetic inter-site correlations are energetically most
favorable. They are followed by rather complex intermediate spin states
(b) as the case of isotropic hopping (c) is approached. Along this
line the salient feature of the correlated ground state is its high
degeneracy. In the case of two dominating hopping channels (d) low-spin
phases with anti-ferromagnetic inter-site correlations are formed.
These features originally derived for periodic boundary conditions
are encountered also for open boundary conditions as can be deduced
from Figure \ref{pd3p}.

\begin{figure}[tb]
\includegraphics[%
  width=0.80\columnwidth]{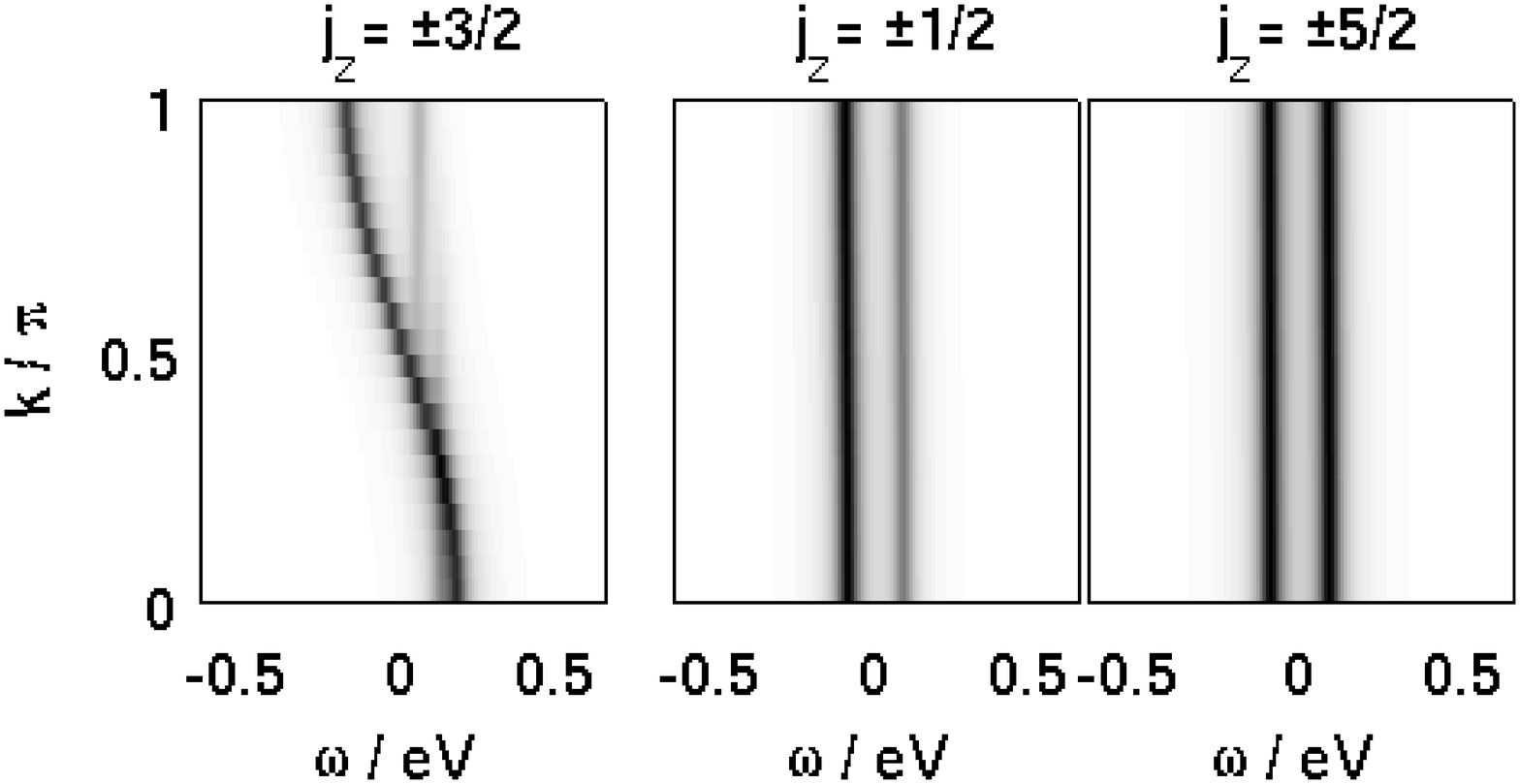}

\includegraphics[%
  width=0.80\columnwidth]{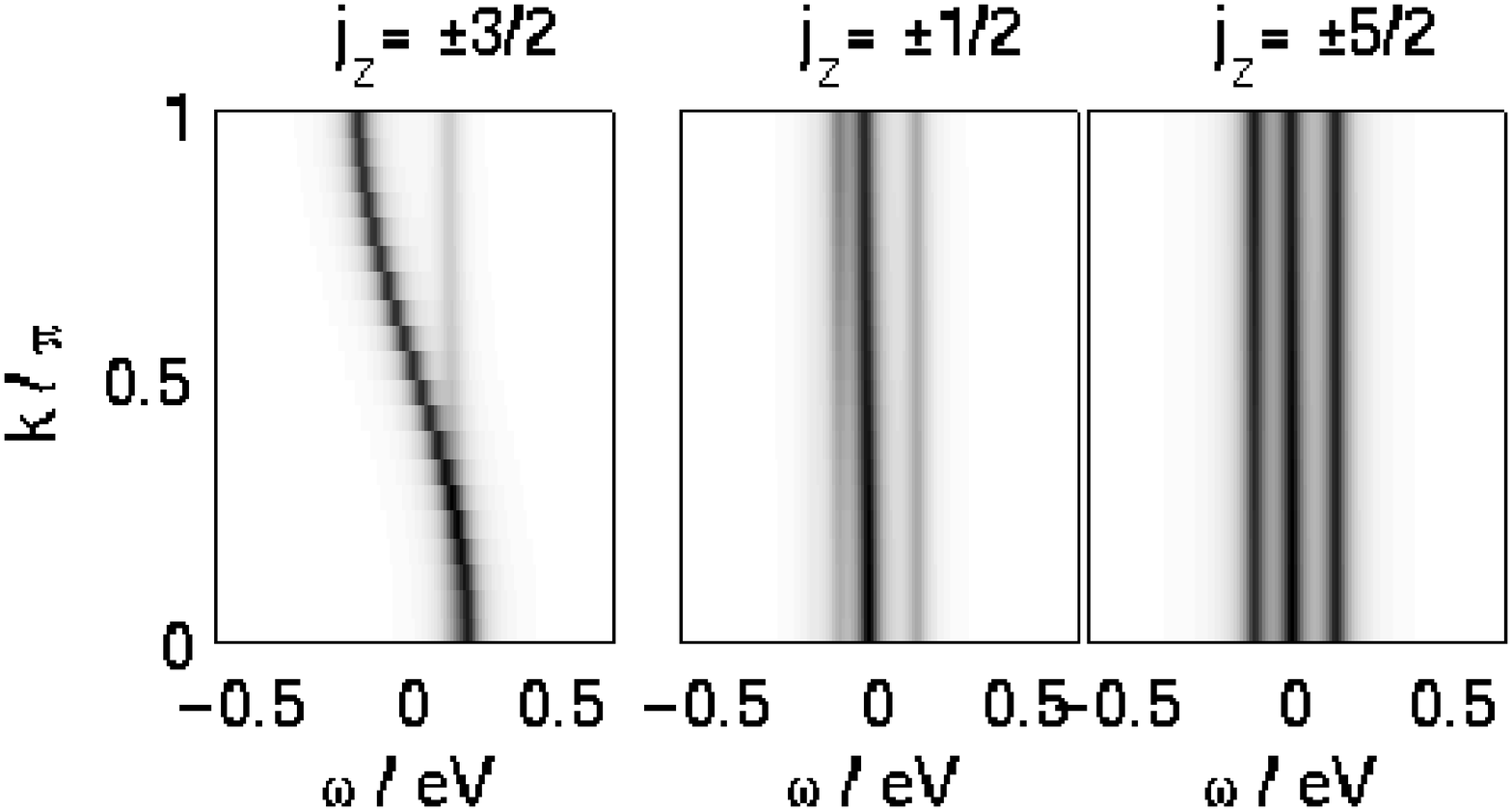}

\caption{Spectral functions of a one-dimensional $5f$-system from the \emph{CPT}
with cluster of two sites (upper panel) and three sites (lower panel).
The orbital-projected spectra are a superposition of the $+j_{z}$
and $-j_{z}$ parts of the spectra. The anisotropic hopping parameters
are and $t_{3/2}=-0.09$ eV and $t_{1/2}=t_{5/2}=-0.01$ eV . The
peaks are broadened with a finite imaginary part of $\eta=0.03$.\label{cap:CPTSpectralfunctionsTwoThreeSitesHighSpinPhase}}
\end{figure}
\begin{figure}[htbp]
\includegraphics[%
  width=0.80\columnwidth]{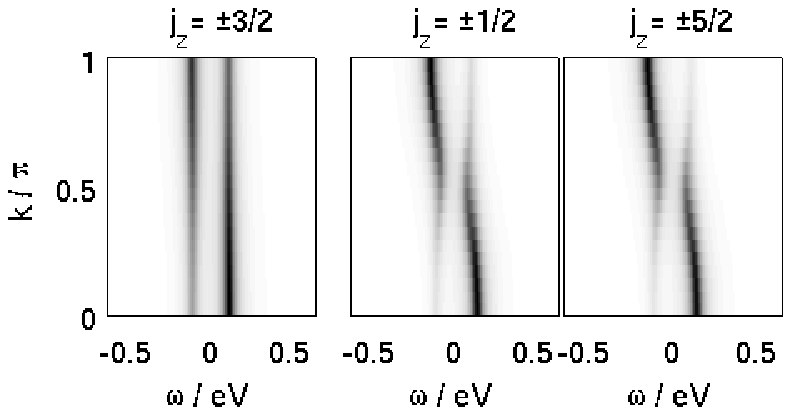}

\includegraphics[%
  width=0.80\columnwidth]{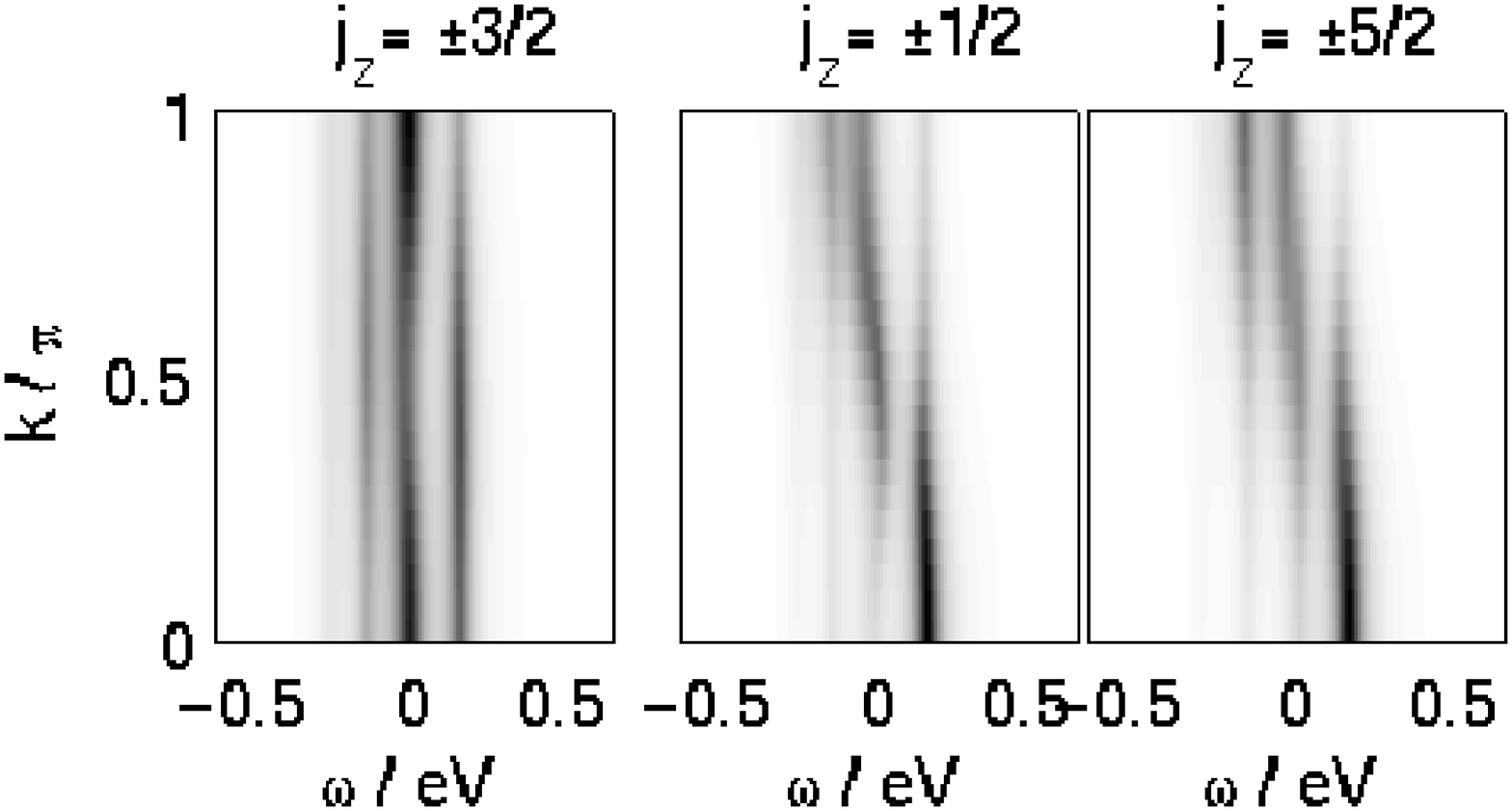}

\caption{Spectral functions of a one-dimensional $5f$-system from the \emph{CPT}
with cluster of two sites (upper panel) and three sites (lower panel).
The orbital-projected spectra are a superposition of the $+j_{z}$
and $-j_{z}$ parts of the spectra. The anisotropic hopping parameters
are and $t_{3/2}=-0.01$ eV and $t_{1/2}=t_{5/2}=-0.09$ eV . The
peaks are broadened with a finite imaginary part of $\eta=0.03$.\label{cap:CPTSpectralfunctionsTwoThreeSitesHighSpinPhase}}
\end{figure}

Figure \ref{cap:CPTSpectralfunctionsTwoThreeSitesHighSpinPhase} displays
the orbital-projected spectral functions of the linear chain. The
data refer to the low-energy regime where the electronic excitations
involve transitions between the lowest Hund's rule multiplets. We
adopt the same values for the transfer integrals as in the cluster
calculations. The system is clearly in the strong-coupling regime
where the symmetries of the ground states are rather sensitive to
the detailed description of the kinetic energy. The high-energy features
associated with transitions to excited atomic states are not shown
here. 

As regards the variation with wave vector of the spectral functions,
the qualitative behavior closely parallels the one obtained for the
finite clusters. The spectral functions clearly show a dispersing
band in the orbital channel with the dominant hopping. The position
of the narrow peak whose finite width is to be attributed to the additional
broadening follows the characteristic cosine dispersion of a one-dimensional
tight-binding band with nearest-neighbor hopping. The gap in the dispersion
of the dominant hopping channel in Figure \ref{cap:CPTSpectralfunctionsTwoThreeSitesHighSpinPhase}
are introduced by the antiferromagnetic intersite correlations which
effectively reduce the size of the Brillouin zone. The band width,
however, is reduced by a factor of $\sim7/12$ as compared to $4\left|t\right|=0.36eV$
expected for uncorrelated electrons. The renormalization reflects
the transfer of spectral weight to the high-energy satellites. The
line widths of the dispersing bands in the linear chain are approximately
twice the value of the bonding-anti-bonding splitting in the two-site
cluster. This discrepancy is due to the fact that the number of nearest
neighbors in the lattice is twice that in the corresponding cluster.
The $5f$ channels with subdominant hopping exhibit low-energy excitations
which are incoherent and which may exhibit a pseudo-gap at the Fermi
energy. It should be noted that the CPT spectra based on three-site
clusters are in good agreement with their two-site cluster counterparts.
The cluster size and the boundary conditions apparently affect the
incoherent excitations in the channels with the subdominant hopping.
This can be seen from comparing the CPT spectra obtained from two-
and three-site clusters.

The co-existence of a coherent $5f$ band and incoherent $f$-derived
low-energy excitations implies partial localization. The orbital-selective
localization suggests the presence of different types of low-energy
excitations which have different orbital character. Photoemission
experiments with polarized incident light should be able to distinguish
the two. In fact, recent experiments on URu$_{2}$Si$_{2}$ seem to
be consistent with this hypothesis.

\begin{figure}[htbp]
\includegraphics[%
  width=0.80\columnwidth]{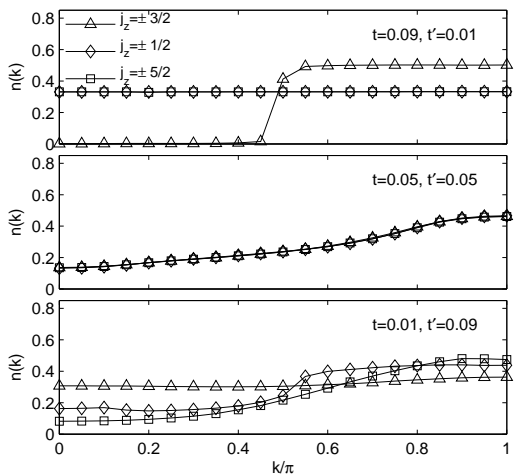}

\caption{Orbital-projected MDF calculated within CPT starting from three-site
cluster and the hopping parameters $t_{3/2}=t$ and $t_{1/2}=t_{5/2}=t'$.
The spectral functions displayed in Figure  were integrated over energy
in an interval of $-4$ eV below the chemical potential.}

\label{cap:OrbitalProjectedMDFCPTThreeSites}
\end{figure}

The Momentum Distribution Function (MDF) as defined in Eq. (\ref{eq:MomentumDistributionFunction})
is displayed in Figure \ref{cap:OrbitalProjectedMDFCPTThreeSites}.
The variation with wave vector $k$ clearly shows the dual character
of the $5f$ electrons. In the homogeneous high-spin phase, the height
of the discontinuity in $n_{j_{z}}(k)$ for orbitals with dominant
hopping is consistent with the spectral weight of the dispersive low-energy
peak. In the low-spin phase, however, it reflects the reduction of
the Brillouin zone due to antiferromagnetic correlations.

\begin{figure}[htbp]
\includegraphics[%
  width=80mm]{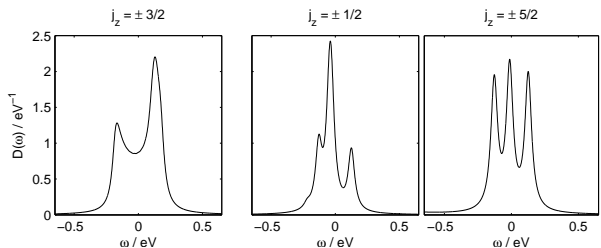}

\caption{Orbital projected density of states $D_{j_{z}}(\omega)$ of a one-dimensional
$5f$-system from the \emph{CPT} with cluster of three sites. The
anisotropic hopping parameters are and $t_{3/2}=-0.09$ eV and $t_{1/2}=t_{5/2}=-0.01$
eV . The peaks are broadened with a finite imaginary part of $\eta=0.03$.\textbf{\label{cap:CPT DOS}}}
\end{figure}
 Figure \ref{cap:CPT DOS} displays the orbital-projected DOS as calculated
within CPT.

\section{Summary and Outlook}

\label{sec:Summary}

We have studied the consequences of strong intra-atomic correlations
on the spectral functions of $5f$ electrons by applying CPT. The
result derived for a one-dimensional chain confirm the idea of orbital-dependent
localization in anisotropic systems. The orbital-projected spectral
functions for the channels with dominant hopping display narrow dispersive
bands in the vicinity of the Fermi energy while the low-energy single-particle
excitations in the remaining channels involve mainly incoherent local
transitions. In all channels, considerable spectral weight is transferred
to the high-energy regime where we find intra-atomic transitions to
excited multiplets as well as valence transitions. The transfer of
spectral weight renormalizes the widths of the dispersing bands and
hence increases the corresponding effective masses.

The dual character of the $5f$ electrons, i. e., the division of
$5f$ states into delocalized and localized orbitals can be clearly
seen form the orbital-projected momentum distribution function. The
latter has a pronounced discontinuity for the orbitals with dominant
hopping while its smooth variation can almost be neglected in the
remaining channels. In a three-dimensional crystal such a behavior
implies that only the orbitals with dominant hopping contribute to
the quantum oscillations of the dHvA effect. 

Calculations starting from clusters with two and three sites yield
similar results in the strong coupling regime. The orbital-dependent
suppression of the kinetic energy calculated for the chain agrees
well with previous results obtained for small finite clusters imposing
periodic boundary conditions. These finding suggest that the features
associated with orbital-selective localization are rather robust. 

The model calculations presented suggest that the orbital-dependence
of the $5f$ spectral functions can be used as fingerprint for orbital-selective
localization in actinide systems. The orbital dependence could be
observable by varying the polarization of the incident light.

It is therefore desirable to perform similar calculations for models
describing real materials. This requires the calculations to be extended
to higher dimensions. An important generalization is the inclusion
of non-diagonal transfer integrals which is currently in progress.

\bibliographystyle{apsrev}
\bibliography{Actinides,ClusterTechniques,HFS,MultiOrbital}

\end{document}